\documentclass[aps,pra,reprint,superscriptaddress,showpacs]{revtex4-1}
\usepackage{graphicx}  
\usepackage{dcolumn}   
\usepackage{pbox}
\usepackage{bm}        
\usepackage{amssymb}   
\usepackage{amsmath}
\usepackage{multirow}

\begin{document}
\renewcommand{\arraystretch}{1.5}
\mathchardef\mhyphen="2D

\title{Measurements of $^{27}$Al$^{+}$ and $^{25}$Mg$^{+}$ magnetic constants for improved ion clock accuracy}

\author{S.~M.~Brewer}
\email{samuel.brewer@nist.gov}
\affiliation{ Time and Frequency Division, National Institute of Standards and Technology, Boulder, CO 80305}
\affiliation{ Department of Physics, University of Colorado, Boulder, CO 80309}
\author{J.-S.~Chen}
\altaffiliation[Current Address: ]{IonQ, Inc., College Park, MD 20740}
\affiliation{ Time and Frequency Division, National Institute of Standards and Technology, Boulder, CO 80305}
\affiliation{ Department of Physics, University of Colorado, Boulder, CO 80309}
\author{K.~Beloy}
\affiliation{ Time and Frequency Division, National Institute of Standards and Technology, Boulder, CO 80305}
\author{A.~M.~Hankin}
\altaffiliation[Current Address: ]{Honeywell Quantum Solutions, Broomfield, CO 80021}
\affiliation{ Time and Frequency Division, National Institute of Standards and Technology, Boulder, CO 80305}
\affiliation{ Department of Physics, University of Colorado, Boulder, CO 80309}
\author{E.~R.~Clements}
\affiliation{ Time and Frequency Division, National Institute of Standards and Technology, Boulder, CO 80305}
\affiliation{ Department of Physics, University of Colorado, Boulder, CO 80309}
\author{C.~W.~Chou}
\affiliation{ Time and Frequency Division, National Institute of Standards and Technology, Boulder, CO 80305}
\author{W.~F.~McGrew}
\affiliation{ Time and Frequency Division, National Institute of Standards and Technology, Boulder, CO 80305}
\affiliation{ Department of Physics, University of Colorado, Boulder, CO 80309}
\author{X.~Zhang}
\affiliation{ Time and Frequency Division, National Institute of Standards and Technology, Boulder, CO 80305}
\affiliation{ Department of Physics, University of Colorado, Boulder, CO 80309}
\author{R.~J.~Fasano}
\affiliation{ Time and Frequency Division, National Institute of Standards and Technology, Boulder, CO 80305}
\affiliation{ Department of Physics, University of Colorado, Boulder, CO 80309}
\author{D.~Nicolodi}
\affiliation{ Time and Frequency Division, National Institute of Standards and Technology, Boulder, CO 80305}
\author{H.~Leopardi}
\affiliation{ Time and Frequency Division, National Institute of Standards and Technology, Boulder, CO 80305}
\affiliation{ Department of Physics, University of Colorado, Boulder, CO 80309}
\author{T.~M.~Fortier}
\affiliation{ Time and Frequency Division, National Institute of Standards and Technology, Boulder, CO 80305}
\author{S.~A.~Diddams}
\affiliation{ Time and Frequency Division, National Institute of Standards and Technology, Boulder, CO 80305}
\affiliation{ Department of Physics, University of Colorado, Boulder, CO 80309}
\author{A.~D.~Ludlow}
\affiliation{ Time and Frequency Division, National Institute of Standards and Technology, Boulder, CO 80305}
\author{D.~J.~Wineland}
\affiliation{ Time and Frequency Division, National Institute of Standards and Technology, Boulder, CO 80305}
\affiliation{ Department of Physics, University of Colorado, Boulder, CO 80309}
\affiliation{ Department of Physics, University of Oregon, Eugene, OR 97403}
\author{D.~R.~Leibrandt}
\affiliation{ Time and Frequency Division, National Institute of Standards and Technology, Boulder, CO 80305}
\affiliation{ Department of Physics, University of Colorado, Boulder, CO 80309}
\author{D.~B.~Hume}
\email{david.hume@nist.gov}
\affiliation{ Time and Frequency Division, National Institute of Standards and Technology, Boulder, CO 80305}

\date{\today}

\begin{abstract}
We have measured the quadratic Zeeman coefficient for the ${^{1}S_{0} \leftrightarrow {^{3}P_{0}}}$ optical clock transition in $^{27}$Al$^{+}$, $C_{2}=-71.944(24)$~MHz/T$^{2}$, and the unperturbed hyperfine splitting of the $^{25}$Mg$^{+}$ $^{2}S_{1/2}$ ground electronic state, $\Delta W / h = 1~788~762~752.85(13)$~Hz, with improved uncertainties.  Both constants are relevant to the evaluation of the $^{27}$Al$^{+}$ quantum-logic clock systematic uncertainty.  The measurement of $C_{2}$ is in agreement with a previous measurement and a new calculation at the $1~\sigma$ level.  The measurement of $\Delta W$ is in good agreement with a recent measurement and differs from a previously published result by approximately $2\sigma$.  With the improved value for $\Delta W$, we deduce an improved value for the nuclear-to-electronic g-factor ratio $g_{I}/g_{J} = 9.299 ~308 ~313(60) \times 10^{-5}$ and the nuclear g-factor for the $^{25}$Mg nucleus $g_{I} = 1.861 ~957 ~82(28) \times 10^{-4}$.  Using the  values of $C_{2}$ and $\Delta W$ presented here, we derive a quadratic Zeeman shift of the $^{27}$Al$^{+}$ quantum-logic clock of $\Delta \nu / \nu = -(9241.8 \pm 3.7) \times 10^{-19}$, for a bias magnetic field of $B \approx 0.12$~mT.
\end{abstract}


\maketitle

\section{Introduction}
Optical atomic clocks based on trapped, laser cooled ions have long been at the forefront of precision frequency metrology \cite{Ludlow2015RMP, ChouAlAlcomparison, Huntemann2016PRL}.  Optical clocks based on ions or neutral atoms have proven to be promising candidates as a replacement for Cs as the definition of the SI second, as well as useful instruments for studies of fundamental physics \cite{Ludlow2015RMP,Rosenband2008Science, Chou2010Science, Godun2014PRL, Huntemann2014PRL}.  An important measure of clock performance is the systematic uncertainty, characterized by the uncertainties associated with all known effects that shift the clock frequency.  The total systematic uncertainty is generally limited by the characterization of the environment in which the clock operates and by uncertainties in the atomic constants needed for evaluating the environment's effect on the atomic resonance frequency.  One environmental factor that must be accounted for is the influence of external magnetic fields on the clock frequency \cite{Gan2018PRA}.  Here we present measurements of magnetic constants, with improved uncertainties, relevant to the systematic uncertainty evaluation of the $^{27}$Al$^{+}$ quantum-logic clock at NIST which employs a single $^{27}$Al$^{+}$ clock ion co-trapped with a single $^{25}$Mg$^{+}$ logic ion.

\begin{figure}[]
\includegraphics[angle=0,width=1.0\columnwidth]{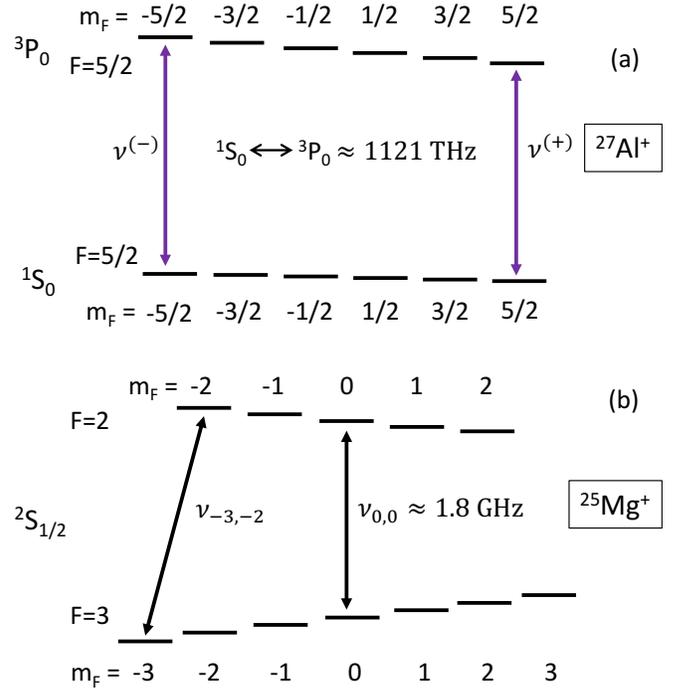}
\caption{\label{fig:elevels} (a) Energy levels and relevant transitions measured for the determination of the $^{27}$Al$^{+}$ quadratic Zeeman coefficient, $C_{2}$.  (b) Energy levels and relevant transitions measured for the determination of the unperturbed $^{25}$Mg$^{+}$ hyperfine splitting, $\Delta W$.}
\end{figure}

The Zeeman structure of the $^{27}$Al$^{+}$ ion, illustrated in Fig.~\ref{fig:elevels}a, consists of six levels $m_F = \{-5/2,\ldots,+5/2\}$ in both the ground $^{1}S_{0}$ and excited $^{3}P_{0}$ electronic clock states.
The energies of these states depend on the external applied magnetic field, $B$, as well as stray magnetic fields that may vary in time. During operation, the clock ideally synthesizes a frequency that is equal to the transition frequency between the ground and excited states at zero magnetic field.  To reach a systematic uncertainty below $10^{-18}$, the Zeeman shift must be accounted for up to 2nd-order in $B$.  The linear Zeeman shift is compensated by interleaved measurements of two transitions: ${|^{1}S_{0}, m_{F} = +5/2\rangle\leftrightarrow|^{3}P_{0}, m_{F} = +5/2\rangle}$ and ${|^{1}S_{0}, m_{F} = -5/2\rangle\leftrightarrow|^{3}P_{0}, m_{F} = -5/2\rangle}$.  The error signal used to correct the laser frequency is generated by taking the average frequency of these two resonances, producing a virtual resonance at the mean of the two frequencies that is first-order insensitive to the magnetic field.  This scheme provides a real-time measure of the static (DC) component of the magnetic field based on the frequency difference between the two transitions.  Given the magnetic field measured from $^{27}$Al$^{+}$ spectra (see below) and the quadratic Zeeman shift coefficient $C_{2}$, a second-order Zeeman correction is applied to recover the unperturbed clock frequency.

In addition to the DC component from the bias magnetic field, there exists an oscillating (AC) magnetic field at the trap drive frequency experienced by the ion primarily due to unbalanced currents in the trap electrodes.  To account for the total magnetic field induced frequency shift on the clock transition, both the DC and AC components are measured.  Since the ion spacing is small ($\approx 5~\mu$m) compared to the ion-electrode distance, the AC component of the field is nearly equal at the location of both ions and is therefore measured using microwave spectroscopy of the co-trapped $^{25}$Mg$^{+}$ ion.  The relevant energy levels and transitions in $^{25}$Mg$^{+}$ are shown in Fig.~\ref{fig:elevels}b.

Historically, the uncertainty in the $^{27}$Al$^{+}$ quadratic Zeeman shift has been limited by the uncertainty in $C_{2}$ and the uncertainty in the $^{25}$Mg$^{+}$ hyperfine splitting, $\Delta W$, and not by the determination of the magnetic field \cite{ChouAlAlcomparison}.  Here, we present measurements of the $^{27}$Al$^{+}$ $C_{2}$ coefficient and the $^{25}$Mg$^{+}$ hyperfine splitting with improved uncertainties compared to previous work.  The measurement of $C_{2}$ is detailed in Sec.~\ref{sec:C2meas} and is compared to a theoretical calculation in Sec.~\ref{sec:C2theory}.  The measurement of $\Delta W$ is presented in Sec.~\ref{sec:Ahfs} and the magnetic field induced $^{27}$Al$^{+}$ clock frequency shift and uncertainty for typical clock operating conditions are discussed in Sec.~\ref{sec:Alunc}.

\section{\label{sec:C2meas}$^{27}$A\lowercase{l$^{+}$} Quadratic Zeeman Shift and $C_{2}$ Coefficient}

The details of the $^{27}$Al$^{+}$ quantum-logic clock at NIST have been presented elsewhere \cite{ChouAlAlcomparison, Chen2017PRL, Chen2017thesis}.  Briefly, a single $^{27}$Al$^{+}$ ion is simultaneously confined with an auxiliary ion in a radiofrequency (RF) Paul trap.  The auxiliary ion is used for sympathetic cooling and readout of the $^{27}$Al$^{+}$ ion internal state using the quantum-logic spectroscopy technique \cite{Schmidt2005Science}.  In the current setup, a $^{25}$Mg$^{+}$ ion is used as the auxiliary ion.

Considering only 1st- and 2nd-order terms in $B$ and ignoring the effect of transverse AC fields coupling adjacent Zeeman levels \cite{Gan2018PRA}, the two atomic resonance frequencies probed by the clock are given by
\begin{equation}\label{eq:freqpm}
\nu^{(+),(-)} = \nu_{0}\pm\frac{5}{2}(g_{p}-g_{s})\frac{\mu_{B}}{h}\langle B\rangle+C_{2}\langle B^{2}\rangle,
\end{equation}
where $\nu_{0}$ is the unperturbed resonance frequency, $g_{s}$ and $g_{p}$ are the g-factors for the ground and excited states, respectively, $C_{2}$ is the coefficient quantifying the quadratic Zeeman shift, $\mu_{B}$ is the Bohr magneton and $h$ is Planck's constant.  Here, brackets $\langle\ldots\rangle$ denote a time average and $+$($-$) refers to the $+5/2$($-5/2$) transition.  The dominant AC components of the magnetic field occur at 60 Hz and harmonics and the trap RF drive frequency (40~MHz and 76~MHz in this work).  Since these frequencies are well below the frequency of the fine structure splitting $(\omega_{10} / 2\pi = (E(^3P_1) - E(^3P_0))/h \approx 1.8~\rm{THz})$, the fractional difference between the AC and DC magnetic polarizability (see Eq.~(\ref{Eq:beta})) is $\lesssim 10^{-9}$ and we neglect any frequency-dependence of the atomic response.  The average magnetic field during clock operation is determined by the difference frequency $\nu^{(+)}- \nu^{(-)}$ and is given by
\begin{equation}\label{eq:Bmean}
\langle B\rangle = \frac{h(\nu^{(+)}- \nu^{(-)})}{5(g_{p}-g_{s})\mu_{B}}.
\end{equation}
The g-factor difference $g_{p}-g_{s} = -1.18437(8)\times10^{-3}$ has been measured by simultaneously measuring $\nu^{(+)}- \nu^{(-)}$ and $\langle B\rangle$ in Ref.~\cite{Rosenband2007PRL}.  Similarly, the coefficient ${C_{2} = -7.1988(48)\times10^{7}}$~Hz/T$^{2}$ has previously been measured by comparing the frequency of the $^{27}$Al$^{+}$ clock transition with a second optical clock while varying the magnetic field~\cite{Rosenband2008Science}.
\begin{figure}[]
\includegraphics[width=1.0\columnwidth]{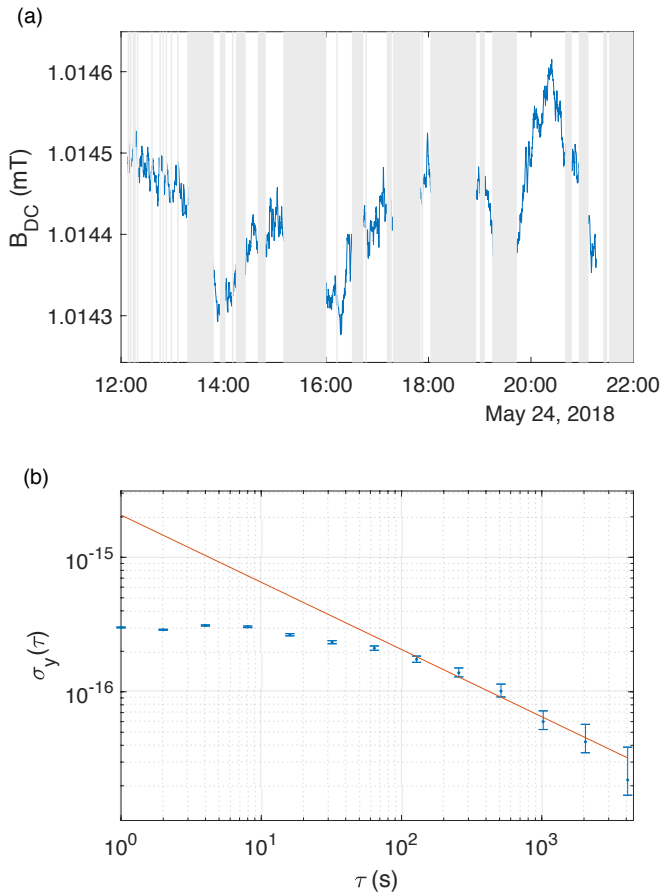}
\caption{\label{fig:Bfieldt} Measurement of the DC magnetic field at the location of the $^{27}$Al$^{+}$ ion.  (a) The magnetic field $B_{DC}$ (See Eq.~\ref{eq:Bmean}) is shown as a function of time.  Sections in gray indicate times when the $^{27}$Al$^{+}$ clock was not operating due to ion loss.  (b) Allan deviation of the $\nu_{\rm Al^{+}} / \nu_{Yb}$ frequency ratio (see text) at $\approx$ 1~mT bias magnetic field in the $^{27}$Al$^{+}$ clock.  The asymptote of the ratio is fit (red line) to a stability of ${\sigma(\tau) = 2.1 \times 10^{-15}/\sqrt{\tau}}$, where $\tau$ is the averaging time in seconds.}
\end{figure}

Ideally the clock servo synthesizes the mean,
\begin{equation}\label{eq:nuion}
\nu_{\rm ion} = \frac{1}{2}\left(\nu^{(+)}+\nu^{(-)}\right) = \nu_{0}+C_{2}\langle B^{2}\rangle,
\end{equation}
which includes the 2nd-order Zeeman shift from both the static applied field and any stray time varying fields.  These two components are treated separately using ${\langle B^{2} \rangle = B_{DC}^{2} + \langle B_{AC}^{2} \rangle}$ such that
\begin{equation}\label{eq:nu0}
\nu_{0} = \nu_{\rm ion}-C_{2} B_{DC}^{2} - C_{2}\langle B_{\rm AC}^{2}\rangle.
\end{equation}
The largest systematic B-field correction to the clock frequency is the 2nd-order Zeeman shift due to the static magnetic field,
\begin{equation}
\nu_{\rm Z,DC} = C_{2} B_{DC}^{2},
\end{equation}
which, at a typical bias field of $B = 0.121$~mT corresponds to a shift of about 1.05 Hz, or $\Delta \nu / \nu \approx 10^{-15}$, expressed fractionally.

Uncertainty in the determination of $\nu_{\rm Z,DC}$ is dominated by uncertainty in the value of $C_{2}$, with
\begin{equation}
\sigma (\nu_{\rm Z,DC}) \approx \sigma_{C_{2}} B_{DC}^{2}.
\end{equation}
For typical clock operation, $\sigma (\nu_{\rm Z,DC}) = 0.7$~mHz in absolute uncertainty. For comparison, uncertainty in the value $g_{p} - g_{s}$ used to determine $B$ contributes about 0.14~mHz and uncertainty associated with statistical fluctuations in the value of the magnetic field is bounded at the same level.

To improve the accuracy of the quadratic Zeeman shift correction, we have measured the frequency shift as a function of the bias magnetic field along the quantization axis in the trap.  The $^{27}$Al$^{+}$ clock bias field was varied from 0.12~mT to 1.01~mT and the frequency was compared to the NIST Yb optical lattice clock frequency.  The NIST Yb lattice clock has a systematic uncertainty of $1.4 \times 10^{-18}$ and a frequency stability characterized by the Allan deviation of $\sigma(\tau) = 1.4 \times 10^{-16}/\sqrt{\tau}$, where $\tau$ is the averaging period of the measurement in seconds.  These and additional details of the Yb lattice clock are presented elsewhere \cite{McGrew2018Nature}.  A pair of octave-spanning frequency combs were used to compare the frequencies of the $^{27}$Al$^{+}$ clock and the Yb clock by locking the repetition rate of the combs to the atom-stabilized Yb clock laser and then counting the beat note frequency between the atom-stabilized $^{27}$Al$^{+}$ clock laser and the nearest frequency comb line.  Measurements of the frequency ratio were made on five different days over the span of two months.  The ratio stability was typically $\sigma(\tau) = 2.1 \times 10^{-15}/\sqrt{\tau}$.  An example of one such real-time measurement of the magnetic field at the location of the $^{27}$Al$^{+}$ ion is shown in Fig.~\ref{fig:Bfieldt}a, with the corresponding frequency stability shown in Fig.~\ref{fig:Bfieldt}b.

\begin{figure}[]
\includegraphics[width=1.0\columnwidth]{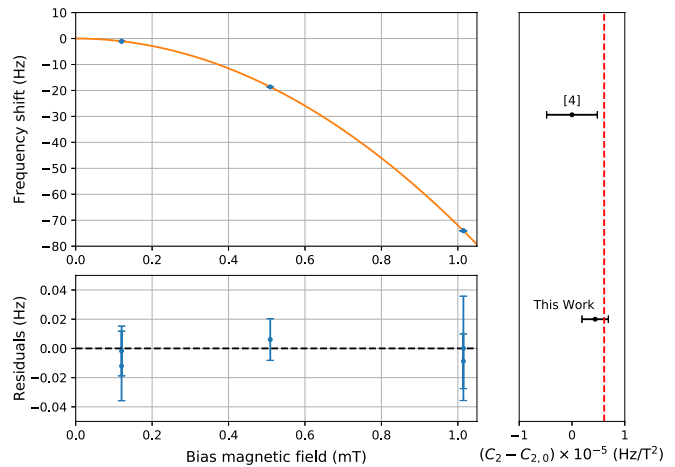}
\caption{\label{fig:ZS2v3} Measurement of the $^{27}$Al$^{+}$ frequency shift as a function of the bias magnetic field.  The frequency shift is fit to a quadratic function to extract the quadratic Zeeman coefficient $C_{2}$.  The fit residuals are shown below.  A comparison to the previous $C_{2}$ measurement ($C_{2,0}$) \cite{Rosenband2008Science}, as well as a theoretical calculation presented in Sec.~\ref{sec:C2theory} (red dashed line) are shown in the right panel.}
\end{figure}

The results of the measurements as a function of the bias magnetic field are shown in Fig.~\ref{fig:ZS2v3}.  A fit to the Zeeman shift data yields a quadratic Zeeman shift coefficient $C_{2} = -7.1944(24) \times 10^{7}$ Hz/T$^{2}$, where the uncertainty in $C_{2}$ is determined using a Gaussian resampling technique and is limited by the statistical uncertainty in the $\nu_{Al^{+}} / \nu_{Yb}$ ratio measurements.  This value is consistent with the previous measurement \cite{Rosenband2008Science} and is in good agreement with a theoretical calculation presented in Sec.~\ref{sec:C2theory}.  This result reduces the uncertainty in $C_{2}$ by approximately a factor of two compared to the previous $\nu_{Al^{+}} / \nu_{Hg^{+}}$ measurement \cite{Rosenband2008Science}.

\section{\label{sec:C2theory}Theoretical calculation of the $^{27}$A\lowercase{l$^{+}$} $C_{2}$ coefficient}
Here, we describe a theoretical calculation of the $^{27}$Al$^{+}$ $C_{2}$ coefficient.  The effects of the hyperfine interaction have been estimated to contribute at a level which is negligible compared to the current level of measurement uncertainty.  Therefore, neglecting hyperfine structure, the frequency dependent magnetic polarizability for a $J=0$ atomic state $\left|n\right\rangle$ is given by
\begin{equation}
\beta\left(\omega\right)=\frac{2}{3\hbar}\sum_{n^\prime\neq n}\left|\left\langle n||\bm{\mu}||n^\prime\right\rangle\right|^2\frac{\omega_{n^\prime n}}{\omega_{n^\prime n}^2-\omega^2},
\label{Eq:beta}
\end{equation}
where $\bm{\mu}$ is the magnetic dipole operator and the ${\omega_{n^\prime n}=\left(E_{n^\prime}-E_n\right)/\hbar}$ are the unperturbed magnetic dipole-allowed transition frequencies \cite{Cowan1981AtomicSpectra}.  The sum over intermediate states excludes summation over the magnetic quantum number.  The AC portion of the $\langle B^{2} \rangle$ is accumulated over Fourier frequencies well-below the atomic transition frequencies.  Consequently, we can apply a static approximation, $\omega \rightarrow 0$, with the level shifts given by
\begin{equation}
\delta E=-\frac{1}{2}\beta\left( 0 \right)\langle B^{2} \rangle.
\end{equation}

In the non-relativistic limit, the eigenstates of the atomic Hamiltonian are $\left|\gamma LSJm_J\right\rangle$, where $\gamma$ specifies the configuration and the remaining angular momentum quantum numbers are all ``good'' quantum numbers. In the non-relativistic limit, we have
\begin{equation}
\bm{\mu}=-g_L\frac{\mu_B}{\hbar}\mathbf{L}-g_S\frac{\mu_B}{\hbar}\mathbf{S},
\end{equation}
where $\mathbf{L}$ and $\mathbf{S}$ are the total orbital and spin angular momentum operators. The $g$-factors here are $g_L=1$ and $g_S=2\left(1+a\right)$, where $a\approx0.00116$ accounts for QED corrections to the electron $g$-factor (anomalous magnetic dipole moment of the electron) \cite{HannekePRL2008}. The reduced matrix elements of the angular momentum operators between these non-relativistic states are
\begin{widetext}
\begin{gather}
\left\langle \gamma LSJ||\mathbf{L}||\gamma^\prime L^\prime S^\prime J^\prime\right\rangle
=\delta_{\gamma,\gamma^\prime}\delta_{L,L^\prime}\delta_{S,S^\prime}(-1)^{L+S+J^\prime+1}\sqrt{\left(2J+1\right)\left(2J^\prime+1\right)}
\left\{
\begin{array}{ccc}
J & 1 & J^\prime \\
L & S & L
\end{array}
\right\}\sqrt{L(L+1)(2L+1)}\hbar,
\\
\left\langle \gamma LSJ||\mathbf{S}||\gamma^\prime L^\prime S^\prime J^\prime\right\rangle
=\delta_{\gamma,\gamma^\prime}\delta_{L,L^\prime}\delta_{S,S^\prime}(-1)^{L+S+J+1}\sqrt{\left(2J+1\right)\left(2J^\prime+1\right)}
\left\{
\begin{array}{ccc}
J & 1 & J^\prime \\
S & L & S
\end{array}
\right\}\sqrt{S(S+1)(2S+1)}\hbar.
\end{gather}
\end{widetext}
Note that $\mathbf{L}$ and $\mathbf{S}$, and therefore $\bm{\mu}$, only mix states of the same fine structure manifold. For the $3s^2\,{^1\!}S_0$ state, the magnetic polarizability is negligible and for the $3s3p\,{^3\!}P_0$ state the only non-vanishing matrix element in Eq.~(\ref{Eq:beta}) is $\left\langle 3s3p\,{^3\!}P_0||\bm{\mu}||3s3p\,{^3\!}P_1\right\rangle$. Specifically, we find
\begin{equation}
\label{Eq:3P03P1}
\left|\left\langle 3s3p\,{^3\!}P_0||\bm{\mu}||3s3p\,{^3\!}P_1\right\rangle\right|
=\sqrt{2}(1+2a)\mu_B.
\end{equation}
This result can be combined with the $3s3p\,{^3P_0}$--$3s3p\,{^3P_1}$ fine structire splitting to obtain the coefficient $C_2$.  We infer the fine structure splitting from spectroscopic measurments in Refs.~\cite{Guggemos2017thesis, Rosenband2007PRL, Itano2007SPIE}, arriving at $\omega_{10}/2\pi = 1.8241180(2)$~THz in the absence of the hyperfine interaction.  The resulting $C_2$ is
\begin{equation}
C_{2} = -71.927~\text{Hz/mT}^2,
\end{equation}
where the result includes the QED correction to the electron $g$-factor.  The theoretical value for the $C_{2}$ coefficient is $\approx 1.3~\sigma$ larger than the measurement in \cite{Rosenband2008Science} and is in agreement with the $C_{2}$ measurement presented here.
\\
\section{\label{sec:Ahfs}AC Quadratic Zeeman Shift and $^{25}$M\lowercase{g$^{+}$} magnetic constants}
The DC component of the field is measured in real-time during the $^{27}$Al$^{+}$ clock operation.  This measurement is insensitive to the AC component because the Rabi spectroscopy probe time is much longer than the inverse of the lowest frequency where there is significant magnetic field noise (60~Hz).  In order to determine $\langle B_{AC}^{2} \rangle$, a separate measurement of the $^{25}$Mg$^{+}$ hyperfine splitting, $\Delta W$, was made.

Here, we describe the measurement of $\Delta W$, with the relevant $^{25}$Mg$^{+}$ energy levels shown in Fig.~\ref{fig:elevels}b.  Interleaved measurements of the $| F = 3, m_{F} = -3 \rangle \leftrightarrow | F = 2, m_{F} = -2 \rangle$ transition frequency ($\nu_{-3,-2}$) and the $| F = 3, m_{F} = 0 \rangle \leftrightarrow | F = 2, m_{F} = 0 \rangle$ transition frequency ($\nu_{0,0}$) are used to extract $B_{DC}^{2}$ and $\langle B^{2} \rangle$, respectively.  A loop antenna located just outside of a viewport of the vacuum chamber was used to drive the various microwave transitions.  The conversion from frequency to magnetic field is made via the Breit-Rabi formula for a $J = 1/2$ system
\begin{equation}
\label{eq: Mgfreq}
\nu_{(m_{F}, m_{F'})} = \frac{E_{F'=2,m_{F'}} - E_{F=3,m_{F}}}{h} ,
\end{equation}
 where
\begin{widetext}
\begin{equation}
\begin{aligned}
\label{eq: BreitRabi}
E_{F = I \pm 1/2, m_{F}} = -\frac{\Delta W}{2(2I + 1)} + \mu_{B}g_{I}m_{F}B \pm \frac{\Delta W}{2}\sqrt{1 + \frac{2m_{F}x}{I+1/2} + x^{2}}.
\end{aligned}
\end{equation}
\end{widetext}
Here, $I$ is the nuclear spin, $g_{I}$ and $g_{J}$ are the nuclear and electronic g-factors, respectively, where $g_{I}$ is in units of the Bohr magneton, and $x = \mu_{B}(g_{J} - g_{I})B/\Delta W$.  To first-order $\Delta W / h = A_{hfs}(I+1/2)$, where $A_{hfs}$ is the hyperfine constant \cite{Ramsey1956MolecularBeams}.

We measure $\nu_{-3,-2}$ and $\nu_{0,0}$ by locking a microwave synthesizer to the transitions.  For each transition, the $\approx 1.8$~GHz probe frequency is square-wave modulated by half the spectroscopic linewidth $\pm \delta \nu/2$ and the difference in transition probability in these two modulation states is used as a frequency discriminator to feed back on the center frequency.  In addition, we monitor the transition probability at the RF center frequency for each transition to ensure we have sufficient contrast in the lineshape throughout the measurement.  The frequency $\nu_{-3,-2}$ was measured using Rabi spectroscopy with a probe time of $\approx 100~\mu$s.  The frequency $\nu_{0,0}$ was measured using Ramsey spectroscopy with $\pi/2$ pulse durations of $\approx 50~\mu s$ and a Ramsey time of $T_{R} = 20$ ms.  Each measurement cycle begins with 1 ms of far-detuned ($\Delta / 2\pi \approx -415$~MHz) laser cooling followed by 500~$\mu$s of Doppler cooling ($\Delta / 2\pi \approx -20$~MHz) on the ${|^2S_{1/2},F=3, m_{F}=-3\rangle} \rightarrow {|^2P_{3/2}, F=4,m_{F}=-4\rangle}$ transition.  The cooling pulses also serve to prepare the ${|^2S_{1/2},F=3, m_{F}=-3\rangle}$ state through optical pumping.  In the case of a $\nu_{-3,-2}$ measurement, after cooling, the $\nu_{-3,-2}$ transition is interrogated followed by resonant fluorescence detection on the ${|^2S_{1/2},F=3, m_{F}=-3\rangle} \rightarrow {|^2P_{3/2}, F=4,m_{F}=-4\rangle}$ transition.  In the case of a $\nu_{0,0}$ measurement, after cooling, the ${|^2S_{1/2},F=2, m_{F}=0\rangle}$ state is prepared using a series of microwave $\pi$-pulses at the transition frequencies $\nu_{-3,-2}$, $\nu_{-2,-2}$, $\nu_{-2,-1}$, $\nu_{-1,-1}$, and $\nu_{-1,0}$ to coherently transfer the population from the ${|^{2}S_{1/2}, F=3, m_{F}=-3\rangle}$ state to the ${|^{2}S_{1/2}, F=2, m_{F}=0\rangle}$ state.  The $\nu_{0,0}$ transition is then interrogated followed by resonant fluorescence detection.

Interleaved measurements of the $\nu_{-3,-2}$ and $\nu_{0,0}$ frequencies were performed over a period of approximately one hour and the uncertainty in $B_{DC}$ is limited by the statistical uncertainty in $\nu_{-3,-2}$.  In addition, we operated interleaved frequency locks on the four auxiliary microwave transitions that are used for state preparation of $|^{2}S_{1/2}, F = 2, m_{F} = 0 \rangle$ during the $\nu_{0,0}$ measurement.  This was done to ensure that ambient magnetic field drifts would not lead to a loss of contrast on the $\nu_{0,0}$ transition.  To eliminate the AC Stark shift on the $\nu_{0,0}$ transition caused by stray Doppler cooling light, we inserted a shutter directly after the output of the UV doubler that blocks the UV light during the experiment.    
 
\begin{figure}
\includegraphics[width=\columnwidth]{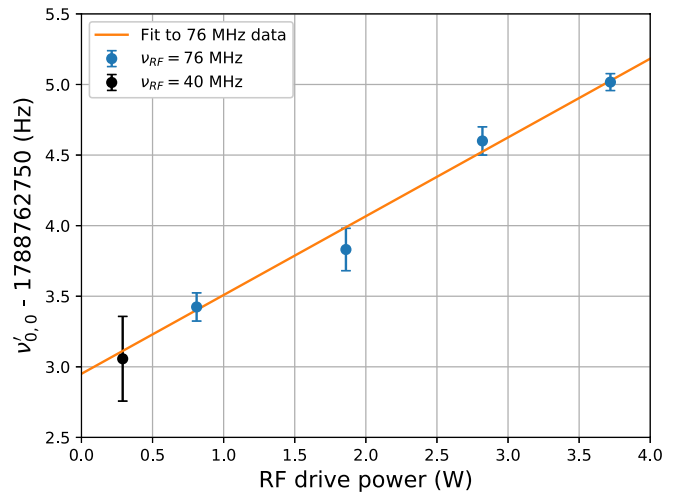}
\caption{\label{fig: F00vsP} Measurement of the $B_{DC}$ corrected $\nu_{0,0}'$ frequency as a function of the trap drive power.  The error bars are statistical and the fit is weighted by those uncertainties.  The data taken at $\nu_{RF} = 76$~MHz was used for the measurement of $\Delta W$ and the data taken at $\nu_{RF} = 40.72$~MHz is used for the evaluation of the AC Zeeman shift in the $^{27}$Al$^{+}$ clock.}
\end{figure}

The frequency, $\nu_{0,0}$ was then corrected for $B_{DC}$ using Eq.~(\ref{eq: BreitRabi}) to give a $B_{DC}$ corrected frequency
\begin{equation}
\nu_{0,0}' = \nu_{0,0} - \frac{\mu_{B}^{2}\left(g_J - g_I\right)^{2}}{2 h \Delta W} B_{DC}^{2}.
\end{equation}
The $\nu_{0,0}'$ transition frequency was measured as a function of the ion trap RF drive power and then fit to a linear function.  The results of these measurements are shown in Fig.~\ref{fig: F00vsP}.  The uncertainty in $\nu_{0,0}'$ is dominated by the statistical uncertainty in $\nu_{0,0}$, whereas uncertainty in the atomic constants based on previous measurements is negligible.  The result of the fit to the measurements performed at an RF drive frequency of 76 MHz, extrapolated to zero RF power, was used to obtain the unperturbed frequency $\Delta W / h$.  The measurement of $\nu_{0,0}'$ at an RF drive power of 0.29~W and a drive frequency of $\Omega_{RF}/2 \pi = 40.72$~MHz is used to evaluate the AC component of the quadratic Zeeman shift and its uncertainty under typical clock operating conditions.  We have measured the magnetic field noise at 60~Hz and harmonics using a set of fluxgate magnetometers positioned just outside the vacuum chamber and estimate that the frequency shift due to these fluctuations is below the statistical uncertainty in $\nu_{0,0}'$.

We have investigated possible additional systematic uncertainties in the $\nu_{0,0}$ measurements.  In particular, there can be an appreciable phase shift on the $\nu_{0,0}$ transition due to the microwave $\pi/2$ pulses used during Ramsey spectroscopy, which can lead to an observed frequency shift.  In the case where the microwave-induced frequency shift $\Delta/ 2 \pi$ is small compared to the Rabi frequency, $|\Delta/\Omega_{0}| \ll 1$, the frequency shift is expressed as \cite{Taichenachev2010JETP},
\begin{equation}
\label{eq: Ramseyshift}
\delta \nu = \frac{\Delta / (2 \pi)}{1 + (\frac{\pi}{4})(\frac{T_{R}}{\tau_{p}})},
\end{equation}
where $T_{R}$ is the free-evolution time and $\tau_{p}$ is the $\pi/2$ pulse duration.  The frequency shift in Eq.~(\ref{eq: Ramseyshift}) leads to an observed frequency $\nu_{0,0} = \nu_{0,0}(T_{R} \rightarrow \infty) + \delta \nu$.  By measuring the frequency shift of $\nu_{0,0}$ as a function of $T_{R}$, assuming the $\Delta W$ value from \cite{Tr2009Ahfs}, we have experimentally determined this shift to be $0.096$~Hz at a probe time of 20 ms.  This effect contributes an additional uncertainty of $\approx 100$~mHz.  This is the dominant systematic uncertainty, leading to a final result of $\Delta W / h = 1~788~762~752.85(13)$~Hz.

\begin{figure}[]
\includegraphics[width=\columnwidth]{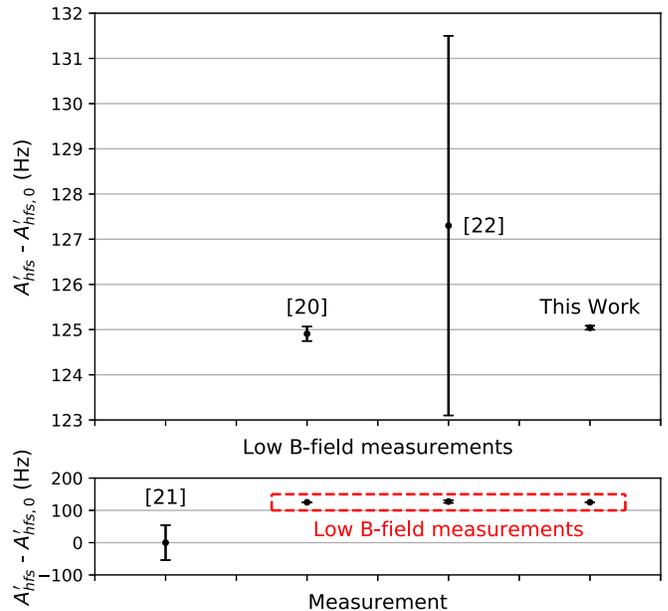}
\caption{\label{fig: Ahfsyr} (Lower panel) Comparison of the uncorrected $^{25}$Mg$^{+}$ hyperfine constant, $A_{hfs}'$, which ignores higher order effects in the hyperfine intraction, from separate experiments.  The $A_{hfs}'$ values are shown relative to the NIST Penning trap measurement, $A_{hfs,0}'$ \cite{Itano1981PRA}.  The  measurements \cite{Tr2009Ahfs,Xu2017PRA} were performed at low magnetic field in an RF trap.  (Upper panel) Low magnetic field based measurements, including the work presented here.}
\end{figure}

Previous work has reported a hyperfine constant $A_{hfs}$ based on measurement of $\Delta W$ \cite{Itano1981PRA,Xu2017PRA}.  At the level of accuracy reported here, however, translating $\Delta W$ to the conventional $A_{hfs}$ constant requires accounting for higher order effects of the hyperfine interaction.  In the ground state of hydrogen, for example, the second order effects contribute at a level of tens of kHz \cite{Latvamaa1973JPhysB}.  Here, for comparison with previous work in Fig.~\ref{fig: Ahfsyr}, we use an uncorrected $A_{hfs}' = -596~254~250.949(45)$~Hz, which ignores any higher order effects.

The $^{25}$Mg$^{+}$ $A_{hfs}'$ value was first measured at NIST in a high-field ($\approx 1$~T) Penning trap \cite{Itano1981PRA}.  The Penning trap results for the hyperfine constant and nuclear-to-electronic g-factor ratio are $A_{hfs}' = -596~254~376(54)$~Hz, and $g_{I}/g_{J} = 9.299~484(75) \times 10^{-5}$.  More recently, $A_{hfs}'$ has been measured in RF traps that employ a low bias magnetic field ($\approx 0.1$~mT) \cite{Tr2009Ahfs, Xu2017PRA}.  In the recent publication by Xu {\it et al.} \cite{Xu2017PRA}, $\nu_{0,0}$ was measured using microwave Rabi spectroscopy.  The result is $A_{hfs}' = -596~254~248.7(4.2)$~Hz, where the uncertainty is dominated by an AC stark shift resulting from residual Doppler cooling light that is not fully extinguished during the frequency measurement.  A summary of $A_{hfs}'$ measurements is shown in Fig.~\ref{fig: Ahfsyr}.  The $A_{hfs}'$ value presented here is in good agreement with both of the results from low-field traps \cite{Tr2009Ahfs, Xu2017PRA}.  All three measurements performed in RF traps agree with each other within the quoted uncertainties and are all $\approx~2\sigma_{\rm combined}$ higher than the previous Penning trap measurement \cite{Itano1981PRA}.

With the improved value of $\Delta W$ and the frequency measurement reported in \cite{Itano1981PRA}, we solve Eq.~(\ref{eq: BreitRabi}) to extract an improved value of the nuclear-to-electronic g-factor ratio $g_{I}/g_{J}~=~9.299~308~313(60)~\times~10^{-5}$.  This result is $\approx 2 \sigma_{\rm combined}$ lower than the previously reported value \cite{Itano1981PRA} and is roughly three orders-of-magnitude more precise.  The uncertainty in $g_{I}/g_{J}$ is dominated by the uncertainty in $\Delta W$.  With the improved determination of $g_{I}/g_{J}$ and $g_{J}$ \cite{Bollinger1992APSBulletin} we deduce a value for the nuclear magnetic moment $g_{I} = 1.861~957~83(28) \times 10^{-4}$.  We caution, however, that for comparison with values of $g_{I}$ derived by other means, careful consideration of diamagnetic corrections may be warranted \cite{Shiga2011PRA}.  For the magnetic field employed ($\approx 0.1$~mT) and resolution of microwave frequencies achieved ($\approx 0.1$~Hz) in our experiment, diamagnetic corrections are expected to be negligible.

\section{\label{sec:Alunc}Quadratic Zeeman shift of the $^{27}$A\lowercase{l}$^{+}$ quantum-logic clock}
The total systematic uncertainty of the $^{27}$Al$^{+}$ quantum-logic clock at NIST has been evaluated and is reported elsewhere \cite{Brewer2019PRL}.  There, the magnetic field induced shifts have been estimated using the improved $C_{2}$ and $\Delta W$ constants presented above.  For typical clock operation, $B_{DC} \approx 0.1208$~mT and the ion trap RF drive frequency is $\Omega_{RF}/2 \pi = 40.72$~MHz.  Under these operating conditions, $\nu_{-3,-2}$ and $\nu_{0,0}$ have been measured to bound $\langle B^{2}_{AC} \rangle$.  As discussed in \cite{Gan2018PRA}, to extract $\langle B^{2}_{AC} \rangle$ from measurements of these hyperfine transition frequencies, we must take into account the direction of the AC field with respect to the quantization axis.  Since our measurement does not distinguish between components of the field parallel to or transverse to the quantization axis, we compute a constraint on the AC magnetic field of ${\langle B^{2}_{AC} \rangle = (1.2 \pm 1.2) \times 10^{-12}~\rm{T}^2}$ which considers a uniform distribution of field directions.  The corresponding clock frequency shift due to the AC component of the magnetic field is ${(\Delta \nu / \nu)_{\langle B_{AC}^2 \rangle} = -(0.8 \pm 0.8) \times 10^{-19}}$.  The DC component of the quadratic Zeeman shift is ${\Delta \nu / \nu_{B_{DC}^2} = -(9241.0 \pm 3.6) \times 10^{-19}}$ and the total shift is ${\Delta \nu / \nu_{\langle B^2 \rangle} = -(9241.8 \pm 3.7) \times 10^{-19}}$.  The uncertainty in the quadratic Zeeman shift is dominated by the uncertainty in $C_{2}$.

\section{Summary and Acknowledgements}
In conclusion, we have presented measurements of magnetic constants relevant to a high-performance $^{27}$Al$^{+}$ optical clock.  The uncertainty in the $^{27}$Al$^{+}$ quadratic Zeeman coefficient $C_{2}$ has been improved by approximately a factor of two, leading to a reduced uncertainty in the quadratic Zeeman shift correction.  The $C_{2}$ value reported here is consistent with a previous measurement \cite{Rosenband2008Science} and agrees with a new theoretical calculation presented here.  Additionally, we have reported a measurement of the hyperfine splitting of the $^{25}$Mg$^{+}$ $^{2}S_{1/2}$ ground electronic state, $\Delta W$, that is consistent with a recent measurement \cite{Xu2017PRA} and has a reduced uncertainty.  From the improved value of $\Delta W$, we also report an improved value of the nuclear-to-electronic g-factor ratio $g_{I}/g_{J}$ and a value for the ion nuclear magnetic moment $g_{I}$.

We thank T. Rosenband and W. Itano for useful discussions, and M.~Kim and M.~Shuker for their careful reading of the manuscript.  This work was supported by the National Institute of Standards and Technology, the Defense Advanced Research Projects Agency, and the Office of Naval Research.  S.M.B. was supported by the U.S. Army Research Office through MURI Grant No. W911NF-11-1-0400.  This article is a contribution of the U.S. Government, not subject to U.S. copyright.

\bibliography{references}
\end{document}